\documentclass[preprint2]{aastex}

\slugcomment{Draft: Do not circulate}

\shorttitle{Nearby Young Solar Analogs}
\shortauthors{Gaidos \& Gonzalez}

\begin{document}


\title{Stellar Atmospheres of Nearby Young Solar Analogs}


\author{Eric J. Gaidos\altaffilmark{1}\altaffilmark{2}}
\affil{Division of Geological \& Planetary Sciences\\
  California Institute of Technology, Pasadena, CA 91125}
\email{gaidos@hawaii.edu}

\and

\author{Guillermo Gonzalez\altaffilmark{3}}
\affil{Astronomy Department, University of Washington, Seattle, WA 98195-1580}
\email{gonzog@iastate.edu}


\altaffiltext{1}{Visiting Astronomer, Kitt Peak National Observatory.
  KPNO is operated by AURA, Inc., under contract to the National Science Foundation.}

\altaffiltext{2}{Current address: Department of Geology \& Geophysics,
University of Hawaii, Honolulu, HI 96822}

\altaffiltext{3}{Current address: Department of Physics \& Astronomy, Iowa State University, Ames, IA 50011}


\begin{abstract}
  High-resolution ($R \approx 90,000$) spectra of 34 nearby, young
  Sun-like stars were analyzed using stellar atmosphere models to
  estimate effective photosphere temperatures, surface gravities, and
  the abundance of certain heavy elements (C, Na, Mg, Si, S, Ca, Ti,
  Fe, and Ni).  The effective temperatures derived from spectroscopy
  were compared with temperatures estimated using optical and
  near-infrared photometry.  In many cases the spectroscopic
  temperatures are significantly higher than the photometric
  estimates, possibly as a result of spottedness or chromospheric
  activity on these active stars.  Values of effective temperature,
  surface gravity, and luminosity were compared to theoretical stellar
  evolution tracks and the evolutionary status of these objects was
  evaluated.  The correlation between heavy element abundance patterns
  and kinematics (space motion) was also examined.  Two nearby stars
  that were tentatively assigned to the Hyades cluster based on
  kinematics have Fe abundances that are also consistent with
  membership in that cluster.  Members of the Ursa Major kinematic
  group exhibit a range of [Fe/H] values but have monotonic [Si/Fe].
  These two observations suggest that heterogeneous incorporation of
  the heavy elements into protostars is creating the variation in
  metallicity.  Local Association members have a distinctly different
  Si/Fe that probably reflects their distinct origin and chemical
  inheritance.

\end{abstract}




\section{Introduction}

Gaidos~(1998) described a catalog of nearby young, solar-type stars
selected as analogs of the Sun during early Earth history.  These
single or wide-binary stars are all within 25 pc and have spectral
types G0-K2 and coronal X-ray luminosities suggesting ages less than
800 My (million years).  Followup spectroscopy and photometery
confirmed marked chromospheric activity, high lithium abundance, and
rapid rotation, all indicative of stellar youth \cite{gaidos00}.

Detailed analysis of the spectra and spectral energy distribution of
stellar photospheres can provide further constraints on the origin and
evolutionary status of these objects.  First, effective temperatures
can be derived by comparing the equivalent widths of iron lines having
different excitation energies with stellar atmosphere model
predictions.  Temperatures can be also estimated from photometry
(i.e., colors) at optical and infrared wavelengths.  Ideally, these
two temperature estimates will be approximately the same (but see
discussion below).  The combination of effective temperature and
luminosity (known by virtue of accurate {\it Hipparcos} parallaxes)
allows the stars to be accurately placed in a Hertzsprung-Russell (HR)
diagram where their locations can be compared with theoretical
zero-age stellar isochrones.  Accurate estimates of effective
temperature are also required for the correct conversion of the
equivalent width of the Li I line at 6708~\AA\ into a lithium
abundance and age proxy.

Second the abundance of heavy elements derived from the equivalent
widths of absorption lines can be used, in principle, as a chemical
``fingerprint'' to relate stars to a common origin, either with other
individual stars or with stellar clusters.  This is potentially useful
because the stars considered here are probably 100-800 My old and no
longer physically associated with molecular clouds or stellar
nurseries.  Solar-mass stars (which do not produce significant amounts
of heavy elements) forming from the same molecular cloud may have
similar metal abundance patterns.  For example, Perryman et al.~(1998)
report [Fe/H] values for 18 members of the nearby Hyades cluster with
effective temperatures between 5000 and 6000 K.  The mean metallicity
is +0.13 with a RMS variation of 0.06, a dispersion consistent with
the measurement accuracy.  Varenne \& Monier~(1999) find a similar
dispersion (but lower mean [Fe/H]) for Hyades F dwarf stars.  In
theory, stars from the same molecular cloud can also retain common
space motions for $\sim 10^9$ yr until gravitational perturbations
from the Galactic tide, molecular clouds, and other passing stars
cause complete kinematic decoherence.  Kinematic clustering is
observed in this catalog \cite{gaidos98,gaidos00} and an objective of
this work is to determine if there is a correlation between space
common space motion and common metal abundance patterns.

\section{Observations and Data Reduction}

High-resolution ($R \sim 90,000$) spectroscopy was carried out at the
Kitt Peak 0.9 m coude feed telescope with the echelle, F5 camera and
3K$\times$1K Ford CCD during four observing runs (May, July, and
November 1998, and March 1999).  Typical total integration times were
$\sim$ 1 hr per star.  Spectral images were bias subtracted and
flattened with quartz lamp flat fields, and charged particle events
were removed by median filtering of multiple exposures.  Individual
spectral orders were extracted, corrected for sky emission, and
wavelength-calibrated against thorium-argon arc spectra using the IRAF
spectroscopic data reduction package.  Equivalent widths were measured
by hand using the IRAF spectrum analysis package.  The atomic data for
the 72 lines used in the analysis are given in Table
\ref{table.lines}.

Near-infrared magnitudes in the $J$, $H$, and $K$ passbands (1.25,
1.65, and 2.2 $\mu$m) measured by the Two Micron All Sky Survey
(2MASS) were extracted from the Infrared Science Archive (IRSA) using
the Gator query engine.  Additional near-infrared measurements were
obtained from the Catalog of Infrared Observations, Version 4.1
\cite{gezari93}.  Most stars also have 12 $\mu$m fluxes reported in
source catalogs generated from observations by the Infrared
Astronomical Satellite ({\it IRAS}) \cite{moshir90}.  Visual (0.55
$\mu$m) magnitudes were obtained from the {\it Hipparcos} catalog
\cite{perryman97}.

\section{Analysis}

\subsection{Spectroscopic analysis}

The method of spectroscopic analysis employed in the present study
follows that employed by Gonzalez et al.~(2001) in their study of the
solar-type parent stars of extrasolar planets.  Briefly, we make use
of equivalent widths (EW's) of relatively unblended Fe I and Fe II
absorption lines with the Kurucz~(1993) plane-parallel model
atmospheres and the LTE abundance code MOOG, originally written by
Sneden~(1973), to determine the four basic atmospheric parameters
(effective temperature, surface gravity, metal abundance, and
microturbulence parameter) for each star. We adopted the atomic data
in Gonzalez et al. and references cited therein. Although that model
was calibrated for a different telescope-spectrograph combination, the
agreement between the parameters of HD 10780 derived from these new
data and those on which the model was originally calibrated suggests
no introduction of significant systematics (see \S 3.3).  The
uncertainty in each parameter was determined using the procedure
described in Gonzalez \& Vanture~(1998). Values of the photosphere
parameters and the number of Fe lines used are listed in
Table~\ref{table.stars}.  The abundances of other elements are also
based on EW measurements and given in Table~\ref{table.abundance}.
The values for HD 1237, a far southern star, are taken from Gonzalez
et al.~(2001).  An acceptable model fit to the data for HD 82443 could
not be obtained: It appears that emission from the chromosphere of
this very active star has significantly filled some photospheric
absorption lines.

\subsection{Photometric Temperature Analysis \label{sec.photometric}}

Astronomical magnitudes were converted to fluxes using an empirical
blackbody formulation of the infrared zero-magnitude fluxes that is
accurate to 1\%.  Weighted least-aquare fits of theoretical
Kurucz~(1992) spectral energy distributions (SEDs) were made to the
data: The two free parameters were the effective temperature $T_{\rm
e}$ and the effective bolometric radius $R$ of the star (the distance
is fixed by the {\it Hipparcos} parallax).  The Kurucz models are
available for effective temperature increments of 250 K; interpolated
models were constructed by a quartic temperature average of two
adjancent models.  (The derived temperature was found to be
insensitive to the exact weighting scheme.)  Most of the
spectroscopically-estimated metallicities are within 0.1 dex of solar,
therefore solar metallicity models were used.  One-sigma (68.3\%
confidence) intervals for $T_{\rm e}$ and $R$ were calculated using
the method of constant $\chi^2$ boundaries as described in Press et
al.~(1986), p.  532.  Note that in some cases only two flux
measurements (typically $V$ and $K$) are available and $\Delta \chi^2$
but not $\chi^2$ is properly defined.  Only a very poor fit (reduced
$\chi^2$ of 10.8) could be achieved for the photometric data for HD
109011: Strassmeier et al.~(2000) report that this object is actually
a spectroscopic binary.  The even poorer fit to the data on HD 220182
may demand a different explanation.

Although we neglect color corrections in our conversion of magnitudes
(integrated over a pass band) to fluxes (at a single wavelength),
these color corrections are relatively minor: We estimated the
magnitude of these corrections by convolving the transmission data
from representative filters (KPNO Harris $V$and 2MASS $K_s$) with the
Kurucz SEDs of G0 and K2 stars, spectral types bracketing those of our
program stars.  In the case of the $K_s$ band filter, a typical
instrument response and atmospheric transmission was included.  While
our photometry is a heterogenous data set with different filters,
detectors, and telescopes, these calculations are still appropriate to
determine the representative magnitude of the color corrections.  The
difference in the color correction between the G0 and K2 stars is
0.006 magnitudes in the $K_s$ bands and 0.067 magnitudes in the
$V$-band.  We estimated the worst-case effect of the latter on the
photometric temperature estimate for a ``typical'' star (HD 20630).
This was done by using {\it only} the $V$- and $K$-band data and
varying the $V$ magnitude by $\pm$0.033.  This produced a variation in
the estimated temperature of $\pm$50 K.  Inclusion of additional data
from other pass-bands into the model fitting will decrease such errors.

\subsection{Comparison with previous results \label{sec.compare}}

Many stars in this study have been previously studied using photometry
and spectroscopy and estimates of their photosphere parameters
published.  By and large, these previous estimates are in good
agreement with our results.  {\bf HD 166:} Our spectroscopic estimate
of $T_{\rm e}$ is significantly higher than a previous value (5255 K)
derived from color indices and several temperature-sensitive lines
\cite{zboril98}.  {\bf HD 1237:} The stellar parameters estimated by
Santos et al.~(2001) are consistent with the values reported here from
Gonzalez et al.~(2001).  {\bf HD 1835:} Cayrel de Strobel et
al.~(1997) (hereafter, C97) report six high-resolution spectroscopy
measurements of atmospheric parameters with $T_{\rm e}$ ranging from
5673 to 5860 K, and Glushneva et al.~(2000) derive 5669 K from JHK
photometry.  Our results (5675 K and 5792 K) are consistent with these
values.  {\bf HD 10780} The spectroscopic and photometric $T_{\rm e}$
values and the metallicity are essentially identical to those derived
from independent data and reported in Feltzing \& Gonzalez~(2001).
The excellent repeatability may be related to the star's relatively
low level of activity: In fact, Gaidos et al.~(2000) concluded that
despite its original selection based on X-ray emission, it is a
solar-age star and we have included it here only for completeness of
the original sample.  {\bf HD 11131:} Our spectroscopic and
photometric temperatures (5700 and 5629 K) are somewhat lower than
those of C97 (5781 K) and Glushneva et al.~(2000) (5645 K).  {\bf HD
20630 ($\kappa$ Cet):} Ottmann et al.~(1998) obtained atmospheric
parameters at three rotational phases which average to $T_{\rm e} =
5680$ K, somewhat lower than our value.  {\bf HD 30495:} Santos et
al. (2001) find higher $T_{\rm e}$, log~$g$, and [Fe/H] values but a
similar microturbulence parameter.  {\bf HD 36435:} We did not observe
this far southerly star but we include the spectroscopic results of
Santos et al.~(2001).  {\bf HD 43162:} Decin et al.~(2000) estimate an
effective temperature of 5593 K from photometry.  Santos et al.~(2001)
found $T_{\rm e} = 5630$, log~$g$ = 4.57, and [Fe/H] = -0.02,
parameter values distinctly different from those obtained here.  {\bf
HD 59967:} Blackwell \& Lynas-Gray~(1998) derive an effective
temperature of 5732$\pm$52 K with the Infrared Flux Method (IRFM),
while our photometric analysis gives a temperature of 5666 K with 68\%
confidence (1~$\sigma$) intervals of 5584-5752 K.  Our spectroscopic
temperature appears to be about 200 K warmer than ther IRTM estimates.
{\bf HD 72905 ($\pi^1$ UMa):} Our values for the four stellar
parameters are identical, to within the uncertainties, with those of
Ottmann et al.~(1998).  {\bf HD 73350:} Our effective temperature and
metallicity are identical to the values estimated by Favata et
al.~(1997).  {\bf HD 128987} Our effective temperature, metallicity,
and microturbulence parameter are consistent with those derived by
Feltzing \& Gustafsson~(1998).  Our value for the surface gravity
(log~$g$ = 4.49) is closer to their ``photometric'' (4.35) rather than
their spectroscopic value (5.00).  {\bf HD 130948:} While our
estimates of effective temperature and surface gravity agree with
those of Chen et al.~(2000), their metallicity ([Fe/H]=-0.20) and
microturbulence parameters (2.0 km sec$^{-1}$) are significantly
different.  Hobbs~(1985) estimates a roughly similar effective
temperature of 5700 K based on photometery, but derives [Fe/H] =
+0.20.  {\bf HD 152391:} Wyse \& Gilmore~(1995) estimate a metallicity
[Me/H] of -0.08 based on $ubvy$ colors, broadly consistent with our
estimate of [Fe/H] = -0.02.  {\bf HD 206860:} Again, Chen et
al.~(2000) derive effective temperatures and surface gravities that
are consistent with our results, but their [Fe/H] (-0.20) is much
lower.  The values of the microturbulence parameter (2.33 km
sec$^{-1}$) that they derive for this star and for HD 130948 are
implausibly high for a dwarf star and may be a consequence of sampling
an inadequate range of line equivalent widths.

\subsection{Comparison of Spectroscopic \& Photometric Temperature Estimates}

Twenty-five stars in our sample have effective temperature estimates
based on both the spectroscopic and photometric methods.  The
spectroscopic estimates are, on average, 118 K higher than the
photometric estimates.  One possible explanation is that there are
systematic errors in one or both of our two analyses.  As shown in \S
\ref{sec.photometric}, the neglect of color corrections in our
photometric temperature estimates cannot produce errors sufficiently
large to explain the discrepancy.  Furthermore, we see no trend of
discrepancy with temperature.  Of the 7 stars with temperature
discrepancies exceeding 200 K, only two (HD 59967 and HD 128987) have
published spectroscopic or photometric/IRFM temperature estimates (\S
\ref{sec.compare}).  Both of these estimates agree, to within the
errors, with our own values.  We also estimated a temperature from
photometry of the active K dwarf $\epsilon$ Eri.  Our estimate agrees
with the spectroscopy-based effective temperature estimates of Drake
\& Smith~(1993) and Santos et al.~(2001), but is curiously higher than
the values derived from narrow-band or broad-band photometry (Table 1
in Drake \& Smith).  Considering the general agreement of our results
with other published results, we conclude there is no {\it a priori}
basis for suspecting problems with our analyses.

\begin{figure}
\plotone{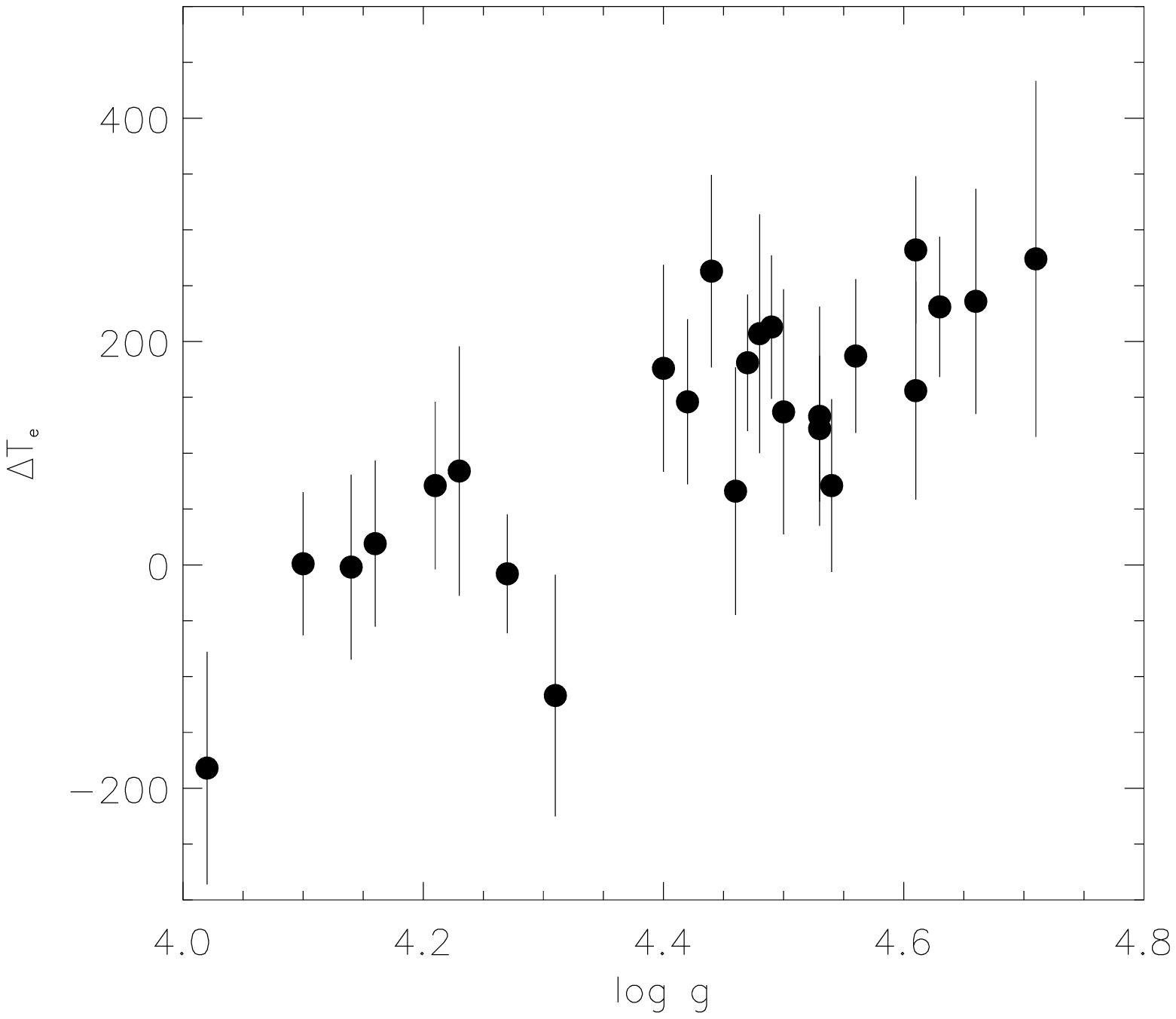}
\caption{Difference between the spectroscopic and photometric
effective temperature estimates vs. surface gravity for the program
star sample.
\label{fig.dtevsg}}
\end{figure}

A second explanation is that the spectroscopic temperature estimates
are corrupted by non-thermal radiation related to stellar activity.
The high surface magnetic fields on these active stars will power
enhanced chromospheric emission and may affect magnetically sensitive
lines such as that of Fe I at 6173 \AA.  Drake \& Smith~(1993) review
evidence that chromospheric activity is responsible for, or at least
associated with, an anomalously large line-to-line scatter in Fe
abundance in solar-type photospheres.  A large scatter could produce
errors in stellar parameters derived from the relative equivalent
widths of lines (e.g., temperature).  Although such scatter is
observed in spectra of the active K star $\epsilon$ Eri, neither Drake
\& Smith~(1993) nor Steenbock \& Holweger~(1981) could find a
correlation between the abundance derived from a particular line and
its Land\'{e} $g$ factor.  Drake \& Smith instead suggest that some
unknown nonthermal excitation or fluorescence effect may be involved.
We find no correlation between the temperature discrepancy and indices
of stellar activity (X-ray luminosity and Ca II HK line intensity).
However, we discovered a significant correlation with surface gravity
(Fig.~\ref{fig.dtevsg}).  The trend of the temperature discrepancy
with log~$g$ is consistent with the expected correlation between
errors in spectroscopic temperature and errors in surface gravity,
suggesting that deviations in the spectroscopic temperature from the
true temperature of the star may play some role.

A third hypothesis is that the photometry is in error or that
photometric estimates are affected by departures of the SED from the
ideal Kurucz models.  Some of the photometric temperature estimates
rely on a $V$-band magnitude in addition to near-infrared photometry.
{\it Hipparcos} visual magnitudes are unlikely to be in error by 0.1
magnitudes or more \cite{soderblom98}.  Bell \& Gustafsson~(1989)
concluded from theoretical model atmospheres that the $V-K$ colors of
dwarf stars with temperatures $\sim$ 5500 K are very weakly dependent
($\sim$ 0.02 mag) on surface gravity and metal abundance.  They also
found that the difference between the temperature estimates based on
Johnson $V-K$ colors and the IRFM have an average of zero and a
standard deviation of only 70 K.

A related hypothesis is that thermal emission from circumstellar dust
increases the total flux from the star at near-infrared wavelengths
and decreases the apparent photometric temperature of the star (the
flux at optical wavelengths could be unaffected).  Gaidos~(1999)
suggested that these stars, as potential young analogs to our Solar
System, may harbor substantial dust clouds.  Very roughly, the K-band
flux of a 5500 K star would have to be augmented by 12\% for its $V-K$
color to be consistent with a 5300 K star.  The dust responsible for
that emission, even at a maximum plausible temperature of 1500 K,
would produce an 80\% excess in the 12 $\mu$m flux relative to the
photosphere.  Such large excess emission has not been observed in data
obtained by the Infrared Astronomical Satellite (IRAS) \cite{gaidos99}
or by the Infrared Space Observatory (ISO) \cite{habing01}.

A fourth hypothesis is that the discrepancy between the two estimates
is real and a consequence of the photometry measuring a
disk-integrated temperature and spectroscopy measuring a value that is
weighted towards the brightest parts of the disk.  Large photospheric
spots, typical on young stars, lower the $V$-band flux and hence the
estimated photometric temperature.  However, the required suppression
of the $V$-band flux seems implausible: The most aggregious case is HD
52698 where the discrepancy between the temperature estimates is 280
K.  Reconciling these two estimates requires that spots dim the
$V$-band magnitude by 0.25 magnitudes (i.e., the $V-K$ color is
reddened by 0.25 magnitudes).  This can be produced by a single
extremely large (r = 30 deg.) spot \cite{amando00}, but the absence of
any modulation of the star's photometric light curve with this
amplitude \cite{gaidos00} requires that the effect would have to be
produced by many, smaller spots.  Furthermore, star spots contain more
intense magnetic fields whose effect is to reduce the gas pressure and
the effective surface gravity in the spots \cite{amando00}.  More
active, spotted stars should appear to have lower surface gravities,
producing a negative correlation between surface gravity and the
temperature difference.  This is opposite to what is observed
(Fig.~\ref{fig.dtevsg}).  Finally, stellar spottedness is also
correlated with stellar activity and a causal link between the
temperature discrepancy and spots would be expected to produce
correlations with stellar activity indices, an effect not seen.

\section{Elemental Abundance Patterns}

\begin{figure}
\plotone{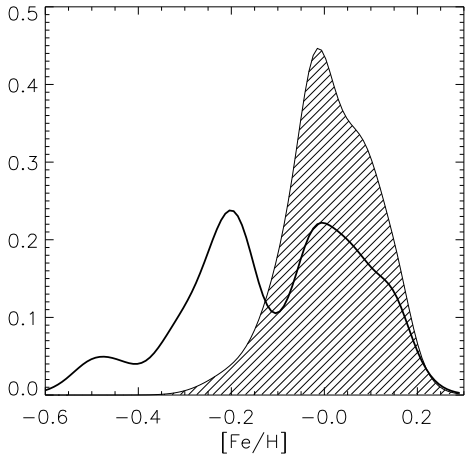}
\caption{Distribution of [Fe/H] among the sample of 33 program stars
(shaded curve).  This is calculated by summing the (normal) error
distribution for each measurement and is in units of the fraction of
the sample per 0.1 dex.  The heavy curve is the [Fe/H] distribution of
the volume-limited sample of Santos et al. (2001)
\label{fig.fehist}}
\end{figure}

{\bf Iron:} The distribution of our spectroscopically-determined
[Fe/H] values for the young stars is plotted in Fig.~\ref{fig.fehist}.
The distribution, with a mean of 0.02 dex and a standard deviation of
0.09 dex, peaks at solar metallicity and is essentially zero at
$\pm0.3$ dex.  In comparison, the [Fe/H] distribution of a
volume-limited sample of 43 solar-mass field stars without detected
planets \cite{santos01} is bimodal and significantly broader
(Fig.~\ref{fig.fehist}).  The average [Fe/H] is -0.11 dex with a
standard deviation of 0.18 dex.  (Similarly, a subsample of 62 G
dwarfs with $T_{\rm e} >$ 5250 K from Favata et al.~(1997) have a mean
of -0.12 and a standard deviation of 0.25) The [Fe/H] distribution of
the young star sample lacks a metal-poor tail and resembles the
metal-rich moiety of the field star distribution.  This is consistent
with the combination of the first sample's narrower range of younger
ages and the age-metallicity relationship in the galactic disk.
Previous studies have also conclusively shown that stars with
detectable planets are statistically more likely to have higher
metallicities than similar ``field'' stars \cite{gonzalez01,santos01}.
Of the three objects that are likely to be the most metal rich (HD
1237, HD 1835, HD 180161) one, HD 1237, is known to harbor a giant
planet \cite{naef01}. Radial-velocity monitoring of HD 1835 places
strict limits on the $m \sin i$ and semi-major axis of any substellar
companions \cite{cumming99}.  While the few available velocity
measurements appear to rule out stellar companions around HD 180161
\cite{tokovinin92,abt94,gaidos98}, high-precision doppler monitoring
of this metal-rich star for planet detection is warranted.

\begin{figure}
\plotone{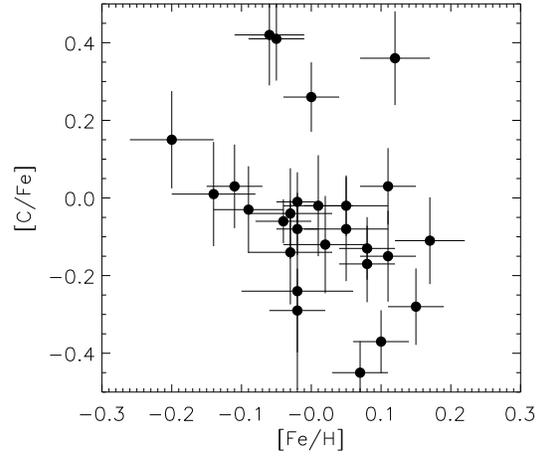}
\caption{Abundance of C relative to Fe in the program star
sample. \label{fig.c}}
\end{figure}

{\bf Carbon:} Carbon abundance is estimated from a single C I line and
consequently has large uncertainties.  The majority of the 29 stars
with C I measurements appear to be carbon-poor relative to the Sun,
with the exception of six stars with [C/Fe] $>$ 0.2
(Fig.~\ref{fig.c}).  These objects are certainly not classical carbon
stars, evolved K and M giants with extremely red ($>$ 2) $B-V$ colors
and [C/O] $\ge$ 1 \cite{ohnaka00}.  They generally do not have high
abundances of the other elements.  More likely, measurement of the
equivalent width of the C I line was affected by nearby weak telluric
(atmospheric water vapor) lines.  The remaining 23 stars exhibit the
well-known trend of decreasing [C/Fe] with increasing [Fe/H]
\cite{gustafsson99,king00a}.  The best linear fit to the 23 points has
a slope of $-0.94\pm0.25$ and a $\chi^2$ of 34.4.

{\bf Sodium:} Values of [Na/Fe] have an average of $-0.12\pm0.01$.
The scatter of the values about the mean is consistent with the
measurement errors ($\chi^2 = 21.6$, $N = 33$), and the data exhibit
no significant trend over the observed range of [Fe/H].  Field stars
with the same (solar) range of [Fe/H] have solar values of [Na/Fe]
\cite{edvardsson93,tomkin95}. The apparent Na-poor character of the
young solar analogs relative to older field stars is contrary to
theoretical expectations: Galactic abundance evolution calculations
suggest that averaged values of [Na/Fe] should increase with time and
hence with [Fe/H] \cite{timmes95}.  Such a trend with [Fe/H] has been
confirmed by surveys of field stars \cite{edvardsson93,feltzing98}.  A
negative trend of [Na/Fe] vs. [Fe/H] described by Edvardsson et
al. 1993 was not confirmed by Chen et al.~(2000).  However, Feltzing
\& Gustafsson~(1998) describe a correlation with kinematics, such that
solar metallicity stars with $Q_{\rm LSR} < 30$ km sec$^{-1}$, where
$Q_{\rm LSR}$ is the total speed with respect to the Local Standard of
Rest, have lower [Na/Fe] than their counterparts with higher space
motions.  Low-speed stars are statistically younger than their
counterparts moving at higher speeds with respect to the LSR and this
suggests that the positive correlation between [Fe/H] and [Na/H] is a
kinematic, rather than chronological, relationship.  All of these
studies use the same Na I 6154/6160 \AA\ line pair.

{\bf Magnesium:} Most [Mg/Fe] values for the young solar analogs are
subsolar, with an average value of $-0.07\pm0.02$.  There is no
significant trend with [Fe/H] and no significant variation from the
mean ($\chi^2 = 14.2$, $N = 32$).  This is in contrast with various
studies of field stars, which generally find [Mg/Fe] $\sim$ +0.1 and
significant scatter \cite{feltzing98}.  Although, a subsolar [Mg/Fe]
in this sample is consistent with the prediction that Fe production by
Type I supernova will depress [Mg/Fe] in younger stars, models of Mg
evolution in the galaxy generally fail to reproduce the observations
\cite{timmes95}.

{\bf Silicon:} [Si/Fe] values are derived from three high-quality
lines (one is a doublet).  The average [Si/Fe] for the sample is
essentially solar ($0.02\pm 0.01$) and the negative trend with [Fe/H]
is of low significance ($-0.13 \pm 0.12$).  (Note the small range of
[Fe/H] spanned by the sample.) This negative trend is seen in both
field stars \cite{edvardsson93,tomkin95} and stars with detected
planets \cite{gonzalez01}.  This trend is most likely a consequence of
Si being the product of massive stars and uncorrelated with Fe
nucleosynthesis and thus the dilution of [Si/Fe] by higher [Fe/H].
The scatter about the mean is comparable to the formal errors ($\chi^2
= 18.3$, N = 33).

{\bf Sulfur:} Sulfur abundance is derived from the equivalent width of
a single, weak line.  The average of the available [S/Fe] measurements
is solar ($0.00\pm 0.03$) and the scatter is consistent with the
measurement errors ($\chi^2 = 35.9$, N = 19).

\begin{figure}
\plotone{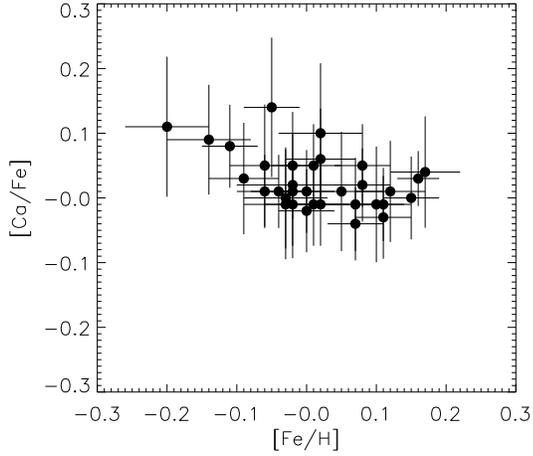}
\caption{Abundance of Ca relative to Fe. \label{fig.ca}}
\end{figure}

{\bf Calcium:} [Ca/Fe] estimates are also derived from high quality
lines.  The average relative abundance is identical to that of Si
($0.02\pm 0.01$).  There is also a negative trend of marginal
significance($-0.16\pm 0.14$) with [Fe/H] (Fig.~\ref{fig.ca}).

{\bf Titanium:} The value of [Ti/Fe] averaged over the sample is
approximately solar ($-0.02 \pm 0.01$) and the individual measurements
exhibit no significant trend with [Fe/H].  The scatter about the mean
is consistent with the measurement errors ($\chi^2 = 24.2$, N = 33).

\begin{figure}
\plotone{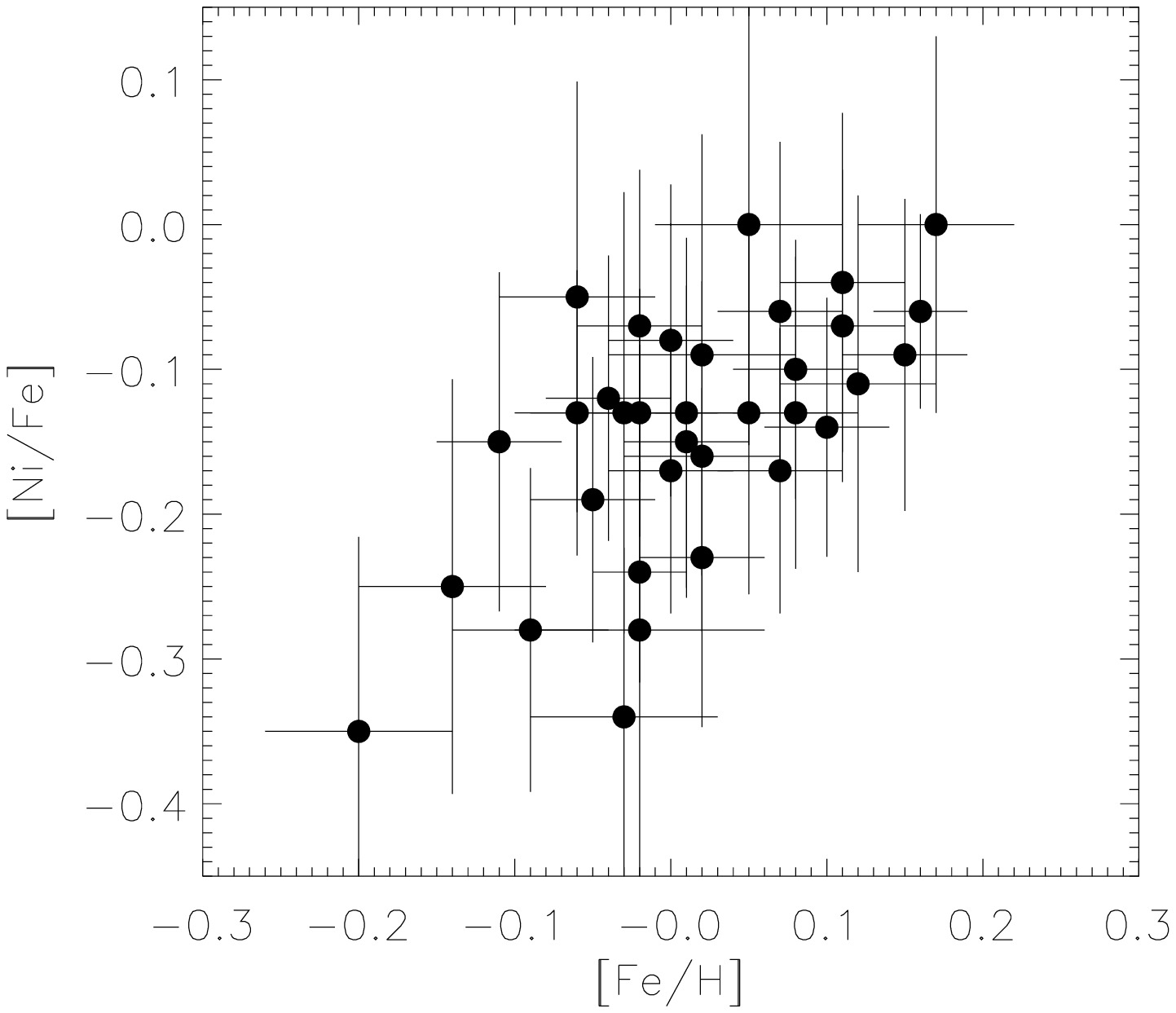}
\caption{Abundance of Ni relative to Fe. \label{fig.ni}}
\end{figure}

{\bf Nickel:} Nickel abundances among the young solar analogs are
decidedly subsolar; (mean value of $-0.15 \pm 0.02$) and exhibit a
strong positive trend with [Fe/H] (slope of $0.60\pm0.22$)
(Fig.~\ref{fig.ni}).  Edvardsson et al.~(1993) report [Ni/Fe] values
with an average of $0.00 \pm 0.03$ over the entire metallicity range
and no significant correlation with [Fe/H].  The isotopes $^{\rm
56}$Ni and $^{\rm 58}$Ni dominate the composition of elemental nickel
and are produced in massive stars and in Type Ia (deflagration)
supernovae.  Galactic abundance evolution calculations \cite{timmes95}
in which most of the Ni is produced in massive stars, are in accord
with the observations of Edvardsson et al.~(1993).  However, the
theory also predicts an overabundance of Ni (relative to solar) by a
factor of 3-4 for the time when and location where the Sun formed.
The positive trend with [Fe/H] observed here and by Gonzalez et
al. suggests that for some stars the contribution by Type Ia
supernovae may be significant.

\section{Discussion}

\subsection{HR Diagram}

\begin{figure}
\plotone{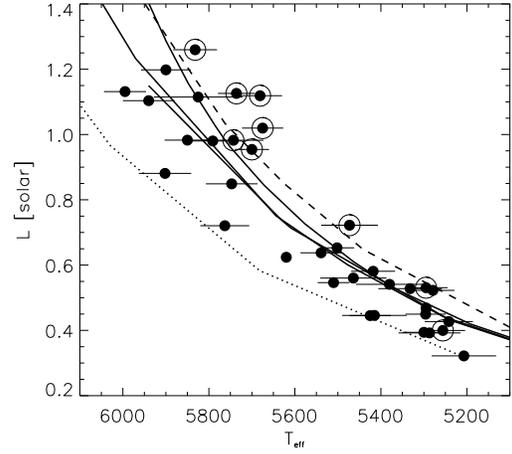} 
\caption{Hertzsprung-Russell diagram (luminosity vs. effective
temperature) for the program stars.  The three solid lines (from lower
to upper) are 0.8-1.2 $M_{\odot}$ solar-metallicity isochrones for
ages of 100, 800, and 4500 million years from Girardi et al.~(2000).
Also plotted are the 800-million year isochrones for $Z=0.008$
(dotted) and $Z=0.03$ (dashed).  The circled points are the stars with
anomalously low surface gravity discussed in the text. \label{fig.hr}}
\end{figure}

The original selection criteria for young solar analog candidates
included the requirement that they lie on or near the Zero Age Main
Sequence (ZAMS) for solar-like metallicities in a Hertzsprung-Russell
(HR) diagram \cite{gaidos98}.  At that time, precision estimates of
photosphere parameters were not available for most candidates and
$B-V$ color was used as a proxy for effective temperature.  Here, we
compare photometrically and spectroscopically derived parameters
(luminosity and effective temperature) with the predictions of stellar
evolution models as a more robust check of these stars' evolutionary
status.  We estimate the luminosity of each star from its absolute
visual magnitude using bolometric corrections derived from a
second-order polynomial fit to the calculated values of Houdashelt et
al.~(2000) over the range 5000 to 6000 K,
\begin{equation}
BC_{\rm V} = -5.30  + 0.00166~T_{\rm e} -1.31\times 10^{-7}~T_{\rm e}^2.
\end{equation}
Fig.~\ref{fig.hr} plots the stars in the HR diagram with the
theoretical models of Girardi et al.~(2000).  

\begin{figure}
\plotone{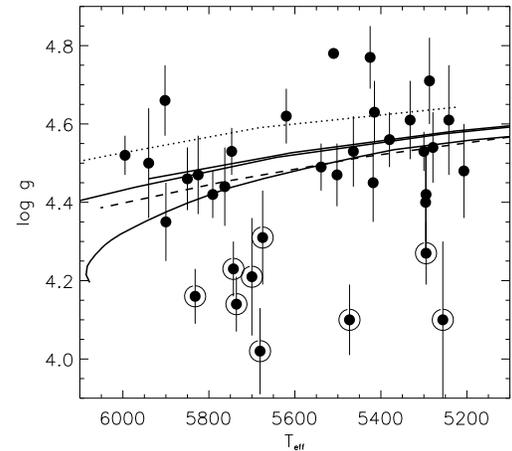}
\caption{Surface gravity vs. effective temperature for the program
stars.  Girardi et al. (2000) isochrones are as in
Fig. 7. \label{fig.g}}
\end{figure}

Although the majority of estimated effective temperatures and
luminosities are consistent with theoretical predictions for young,
solar-metallicity, solar-mass stars there are eight stars (HD 1835, HD
10780, HD 11131, HD 43162, HD 73350, HD 97334, HD 130948, and HD
165185) whose measured luminosities and temperatures displace them
from the solar-metallicity zero-age main seqence.  A high-metallicity
explanation is inconsistent with the spectroscopy-based abundance
values presented here (-0.14 to +0.17 dex).  In addition, these
objects have anomalously low estimated surface gravities with log~$g$
$\sim4.1$ (Fig.~\ref{fig.g}).  HD 118972 also has a low log~$g$ but
lies on the main sequence: The large errors in the surface gravity
indicate a poor fit to the stellar atmosphere model.  HD 10780 has
already been identified as an old star by virtue of its slow rotation
compared to the rest of the sample (Gaidos et al.~2000).  However, the
remaining seven suspect stars are enigmatic: They display all the
relevant characteristics of stellar youth, including high coronal
X-ray luminosity, high chromospheric emission in the Ca II H and K
lines, rapid rotation, and high lithium abundance \cite{gaidos00}.
Two stars (HD 73350 and 97334) are putative members of the Hyades
cluster and the Local Association, respectively (see \S
\ref{kinematics}).

\begin{figure}
\plotone{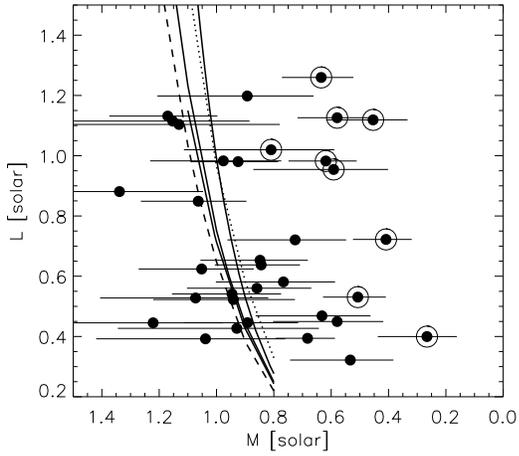}
\caption{Stellar luminosity (derived from absolute magnitude and a
bolometric correction) vs. stellar mass (derived from effective
temperature, surface gravity, and luminosity) of the program stars
expressed in solar units.  Girardi et al.  isochrones are plotted as
in Fig. 7.  The circled stars are nearly all inconsistent with the
model. \label{fig.m}}
\end{figure}

The mass of a star can be uniquely derived from the effective
temperature $T$, bolometric luminosity $L$, and surface gravity $g$
using the relationship
\begin{equation}
\label{eqn.mass}
M = \frac{gL}{4\pi G \sigma T^4},
\end{equation}
where $G$ and $\sigma$ are the gravitational and Stefan-Boltzmann
constants.  This can be re-expressed in terms of the parallax $\pi$
and visual magnitude V \cite{fuhrmann98}:
\begin{equation}
\log \pi = 0.5([g]-[M]) - 2[T_{\rm e}] - 0.2(V + BC_{\rm V} + 0.26),
\end{equation}
where $[X] = \log \left(X/X_{\odot}\right)$.  However, stellar
evolution theory predicts that only some combination of $M$, $L$,
$T_e$, and log~$g$ are possible.  Theoretical evolution tracks can be
used to check the consistency of these parameters \cite{allende99b}.
Fig.~\ref{fig.m} compares derived values of mass-luminosity pairs with
Girardi et al. isochrones.  With the exception of HD 1835, the derived
mass and luminosity values of suspect stars are inconsistent with
their main-sequence status.  Stellar luminosity is estimated from
accurate {\it Hipparcos} astrometric and photometry combined with a
small bolometric correction. Instead, our $T_e$ and/or log~$g$
estimates for these stars appear to be problematic.

Allende-Prieto \& Lambert (1999) estimated atmosphere parameters for
fifteen of our program stars using Eqn. \ref{eqn.mass} and stellar
model predictions: Our measurements of log~$g$ and effective
temperature and their estimates are discrepant by about 0.3 dex and
100 K, respectively.  Error covariance between $T_e$ and log~$g$ might
then explain the trend seen in Fig. \ref{fig.dtevsg}.  One possibility
is that the chromospheric activity has effected certain Fe lines and
corrupted estimates of surface gravity.  However, we found no
significant correlation between stellar activity level and log~$g$
values.  Paradoxically, the photometric estimates of effective
temperature for these stars are in agreement with the
spectroscopy-based results.  We note that of three observations of HD
1237 on successive nights using the same telescope, instrument,
observer, and data reduction procedure, the third produced an
anomalously low log~$g$ estimate but a consistent temperature,
metallicity, and microturbulence parameter \cite{gonzalez01}.

\subsection{Microturbulence}

\begin{figure}
\plotone{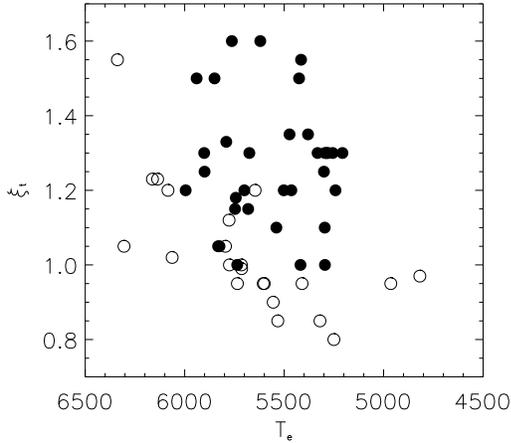}
\caption{Derived microturbulence parameter $\xi_t$ (in km sec$^{-1}$)
vs. effective temperature for the young solar analog stars (filled
points) compared to a sample of older solar-type stars with planets
analyzed in Gonzalez et al. (2001) (empty points).  The $\xi_t$ values
for the young stars are systematically higher for the same effective
temperature.
\label{fig.microturb}}
\end{figure}

The microturbulence parameter $\xi_t$ for all of these stars is equal
to or exceeds 1 km sec$^{-1}$.  The sample mean is 1.30 km sec$^{-1}$
with a standard deviation of 0.15 km sec$^{-}$.  These values are
distinctly higher than those obtained for older G-type stars by
Gonzalez et al.~(2001), using the identical analysis procedures
(Fig.~\ref{fig.microturb}).  The enhanced {\it macroturbulence}
velocity dispersions of magnetically active stars were noted by Saar
\& Osten (1997).  We compared our values with the predictions of the
empirical relation derived by Feltzing \& Gustafsson~(1998) from
measurements of 12 metal-rich G and K dwarf stars;
\begin{equation}
\xi_t = 4.5\times 10^{-4} T_{\rm e} - 0.31 {\rm log}~g.
\end{equation}
The actual values exceed the predicted values by an average of only
0.14 km sec$^{-1}$ and there is significant scatter (reduced $\chi^2$
of 4.3).  While this latter comparison suggests that the measured
microturbulence parameters for these stars are not too abnormal, we
caution that differences in the derivation procedure may limit the
validity of this conclusion.

\subsection{Abundances and Kinematics \label{kinematics}}

Some of these young stars have been previously related to open
clusters or co-moving groups via common space motions \cite{gaidos00}.
One of the objectives of our abundance measurements was to determine
if there is a link between association by space motion and association
by elemental abundances.  Two stars (HD 73350 and HD 180161) were
identified as potential members of the Hyades cluster based on space
motions \cite{gaidos00}.  The [Fe/H] values reported here are also
consistent with such membership (Fig.~\ref{fig.hyades}).  Furthermore,
the solar values of [Si/Fe] for these two stars agree with
measurements on late F-type stars in the Hyades \cite{varenne99}.
This concordance allows us to assign a Hyades age ($\sim$ 650 Myr ;
Perryman et al. 1998) to HD 73350 and HD 180161 with greater
confidence.  The latter star, at a distance of 20 pc, would be the
closest known solar-mass member of the Hyades cluster.

The seven candidate members of the Ursa Major kinematic group
\cite{gaidos00} exhibit a wider range of [Fe/H] values that are still
consistent with previous estimates of the metallicity distribution
\cite{soderblom93} (Fig.~\ref{fig.uma}).  A possible exception is HD
41593, which has an [Fe/H] of $0.06 \pm 0.04$.  The dispersion of the
stars about the Soderblom \& Mayor average [Fe/H] of -0.08 is 0.10
dex, consistent with the previously estimated dispersion and
significantly larger than the measurement errors ($\chi^2 =27.5$, N =
5).  Thus there is {\it significant} variation in the [Fe/H] values
within the Ursa Majoris kinematic group.  If these stars are indeed
coeval and originate from the same molecular cloud, then that cloud of
origin must either have been heterogeneous with respect to Fe
abundance (e.g., as a result of injection of material from previous
generations of forming stars) or that Fe segregated out with varying
degrees of efficiency in the cloud.

In principle, these two mechanisms can be distinguished by examining
the distribution of [Si/Fe] values.  Silicon and iron are contributed
by different mass ranges of stars and it seems implausible that
inhomogeneities produced by stellar mass loss would occur such that
variations in [Fe/H] would not be accompanied by variations in
[Si/Fe].  On the other hand, local variation in abundance that does
not discriminate between Si and Fe (i.e., grain removal by radiation
pressure) would produce constant [Si/Fe] in the resulting stellar
population.  The average [Si/Fe] for these seven stars is +0.02 and
there is no significant dispersion about this mean.  This observation
supports the second mechanism.

Six stars have space motions consistent with membership in a second
kinematic group known as the ``Local Association'' \cite{gaidos00}.
(Note that HD 10008 was previously omitted because of a computational
error).  The existence of the Local Association was first suggested by
Eggen~(1983) and later re-identified in surveys of active and
lithium-rich stars \cite{jeffries93,jeffries95}.  Although a link has
been suggested with the Pleiades cluster, the relative motion between
the Pleiades cluster and the Local Association is significant; 8.8 km
sec$^{-1}$ \cite{gaidos00}.  The age-related properties of Local
Association stars are also inconsistent with a Pleiades age of $\sim$
100 Myr \cite{gaidos00}.  However, the Gaidos sample includes two
young solar analogs (HD 82443 and HD 113449) with space motions
similar to that of the Pleiades.  HD 82443 (DX Leo) has already been
suggested as a nearby Pleiades member \cite{soderblom87} and its
chromospheric activity and X-ray luminosity is suggestive of a very
young age.  Unfortunately, line filling by chromospheric emission
prevented accurate determination of the photosphere parameters or
elemental abundances of this star.

The Fe abundances derived for the six previously-identified Local
Association members span the range -0.02 to +0.11 while HD 113449 has
[Fe/H] = $-0.06\pm 0.05$.  Values of [Na/Fe], [Mg/Fe], [Si/Fe],
[Ca/Fe], [Ti/Fe], and [Ni,Fe] averaged over the Local Association
members are -0.10, -0.11, -0.02, +0.01, -0.02, and -0.14,
respectively.  In particular, the Local Association stars appear to
occupy a distinct part of a [Si/Fe] vs. [Fe/H] plot
(Fig.~\ref{fig.local}), although not all of the stars in that region
are members.  With the possible exception of Ni, the abundance pattern
of HD 113449 is statistically indistinguishable from that of the Local
Association average.  In particular, both have low Na and Mg
abundances relative to Fe.  The difference between the mean space
motion of the Local Association and HD 113449 is 11 km sec$^{-1}$:
However most of this is the difference in velocities along the
relatively unconstrained axis perpendicular to the galactic plane.
The chemical and kinematic similarity suggests a genetic relationship
between HD 113449 and the members of the Local Association.

Boesgaard \& Friel~(1990) obtained an average [Fe/H] of -0.02 for 12
Pleiades F stars, Cayrel et al.~(1985) found [Fe/H] = +0.13 for 4 G
dwarfs, and King et al.~(2000) estimated an average of [Fe/H] = 0.06
for two cool (late K) dwarfs. This broad range of metallicities is
consistent with the Local Association members although HD 113449 is
relatively metal poor.  Detailed abundance data for Pleiades stars is
scarce; the average of the two cool dwarfs analyzed by King et
al. show an abundance pattern that is strikingly different, with solar
[Mg] but depletion of Ca and Ti.  However, it is not known if these
stars are typical and a proper comparison with the Pleiades cluster
will require abundance data for many more {\it bona fide} members.

{\acknowledgments

We are grateful to the personnel of KPNO, especially D. Willmarth, for
courteous and professional assistance in the spectroscopic
observations.  We also acknowledge the constructive comments of two
anonymous reviewers.  EJG was supported by the NASA Astrobiology
Program and GG was supported, in part, by a grant from the Kenilworth
Fund of the New York Community Trust.  This publication makes use of
data products from the Two Micron All Sky Survey, which is a joint
project of the University of Massachusetts and the Infrared Processing
and Analysis Center/California Institute of Technology, funded by the
National Aeronautics and Space Administration and the National Science
Foundation.  The SIMBAD database, maintained by the Centre de
Donn\'{e}es astronomiques de Strasbourg (CDS), and NASA's Astrophysics
Data Systems Bibliographic Services were used extensively.
}

\clearpage


\begin{figure}
\plotone{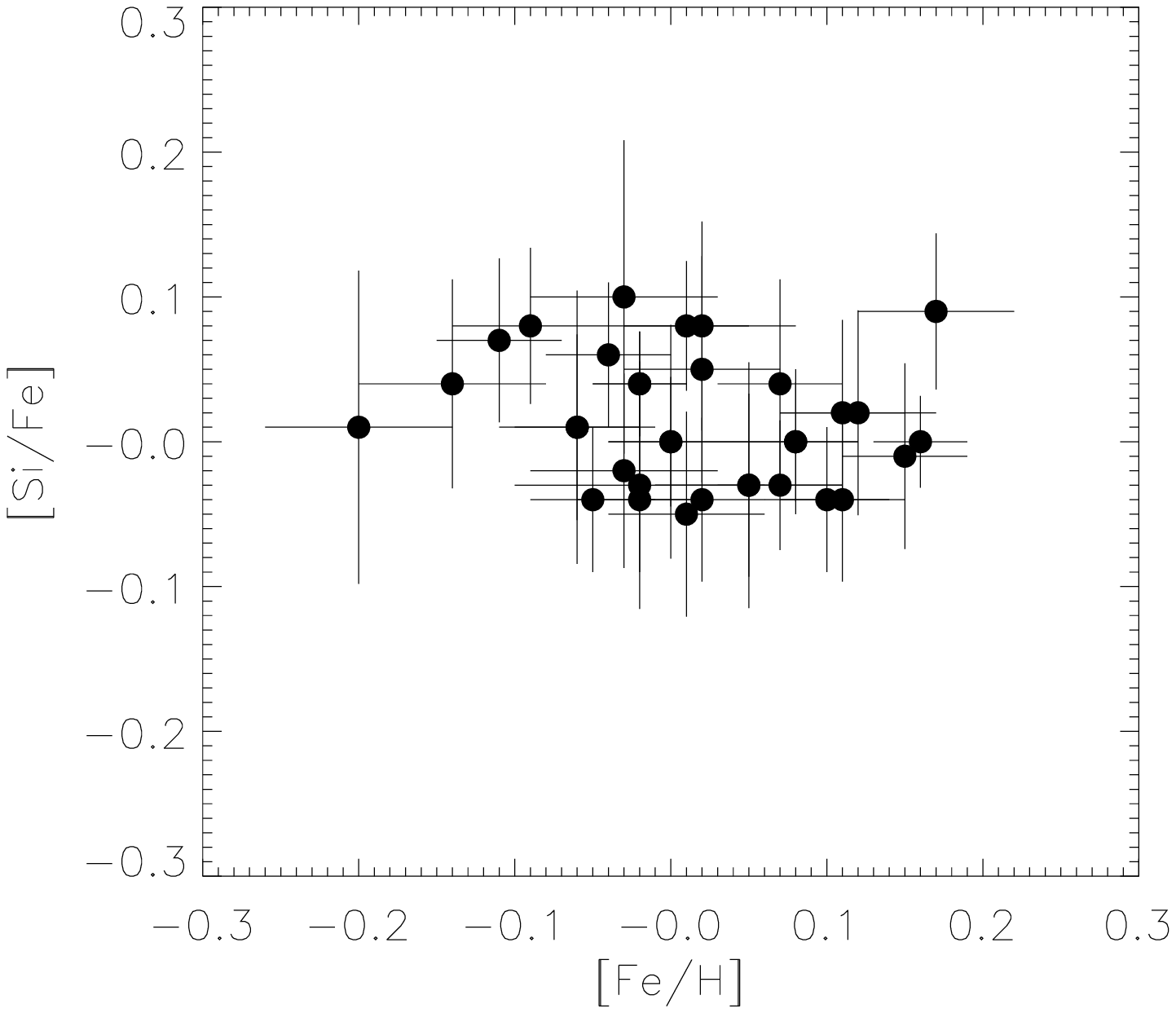}
\caption{Abundance of Si relative to Fe. \label{fig.si}}
\end{figure}

\begin{figure}
\plotone{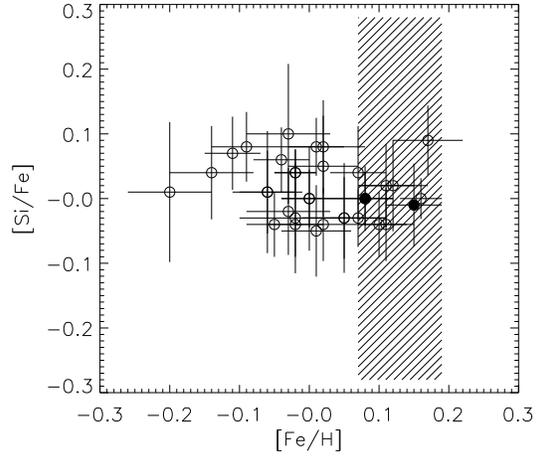}
\caption{Same as Fig.~\ref{fig.si}, except that two Hyades cluster
candidate members are plotted as filled circles.  The hatched region
is the range of [Fe/H] observed in the Hyades cluster within one
standard deviation of the mean. \label{fig.hyades}}
\end{figure}

\begin{figure}
\plotone{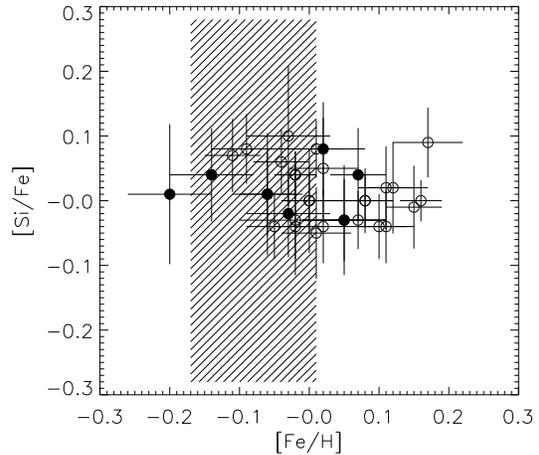}
\caption{Same as Fig.~\ref{fig.si}, except that seven Ursa Major
kinematic group members are plotted as filled circles.  The hatched
region is the range of the group metallicity within one standard
deviation of the mean \cite{soderblom93}. \label{fig.uma}}
\end{figure}

\begin{figure}
\plotone{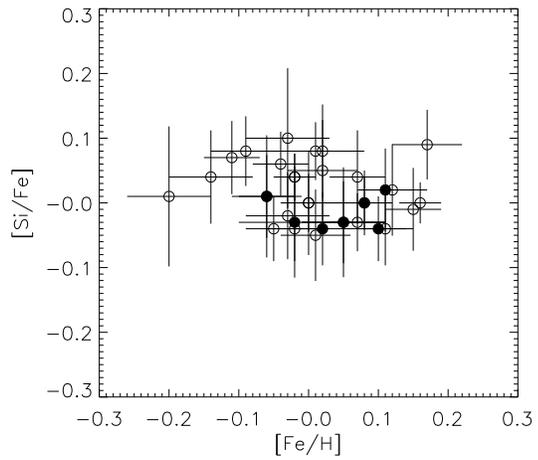}
\caption{Same as Fig.~\ref{fig.si}, except that six Local
Association kinematic group members and HD 113449 are plotted as
filled circles (HD 113449 is the far left point).  \label{fig.local}}
\end{figure}

\clearpage 







\clearpage

\begin{deluxetable}{lrrr}
\tabletypesize{\scriptsize}
\tablecaption{ATOMIC DATA FOR LINES \label{table.lines}}
\tablewidth{0pt}
\tablehead{
\colhead{SPECIES ($\epsilon_{\odot}$)} & \colhead{$\lambda_0$~($\AA$)} & \colhead{$\chi_1$~(eV)} &
\colhead{log~$gf$}
}
\startdata
C~I (8.56).....
& 6587.620  &  8.53  &  -1.08 \\
Na~I (6.33)....
& 6154.230  &  2.10  &  -1.58 \\
& 6160.750  &  2.10  &  -1.26 \\
Mg~I (7.58)....
& 5711.100  &  4.34  &  -1.71 \\
Si~I (7.55)....
& 6125.030  &  5.61  &  -1.54 \\
& 6145.020  &  5.61  &  -1.42 \\
& 6721.840  &  5.86  &  -1.14 \\
& 6721.850  &  5.86  &  -1.14 \\
S~I (7.21).....
& 6052.680  &  7.87  &  -0.44 \\
Ca~I (6.36)....
& 5867.570  &  2.93  &  -1.62 \\
& 5867.572  &  2.93  &  -1.62 \\
& 6166.440  &  2.52  &  -1.13 \\
Ti~I (4.99)....
& 5965.835  &  1.88  &  -0.38 \\
& 5965.840  &  1.88  &  -0.38 \\
& 6126.220  &  1.07  &  -1.41 \\
& 6261.110  &  1.43  &  -0.46 \\
& 6261.220  &  1.43  &  -0.46 \\
Fe~I (0.00)....
& 5852.220  &  4.55  &  -1.18 \\
& 5853.160  &  1.48  &  -5.18 \\
& 5855.090  &  4.61  &  -1.52 \\
& 5856.100  &  4.29  &  -1.56 \\
& 5956.700  &  0.86  &  -4.55 \\
& 6027.060  &  4.07  &  -1.09 \\
& 6034.090  &  4.31  &  -2.26 \\
& 6056.010  &  4.73  &  -0.40 \\
& 6079.070  &  4.65  &  -1.02 \\
& 6089.570  &  5.02  &  -0.86 \\
& 6093.700  &  4.61  &  -1.34 \\
& 6096.670  &  3.98  &  -1.81 \\
& 6098.310  &  4.56  &  -1.74 \\
& 6120.250  &  0.91  &  -5.88 \\
& 6151.620  &  2.18  &  -3.29 \\
& 6157.790  &  4.07  &  -1.25 \\
& 6159.440  &  4.61  &  -1.87 \\
& 6165.360  &  4.14  &  -1.47 \\
& 6180.260  &  2.73  &  -2.61 \\
& 6188.050  &  3.94  &  -1.61 \\
& 6200.320  &  2.61  &  -2.44 \\
& 6226.740  &  3.88  &  -2.03 \\
& 6229.230  &  2.84  &  -2.82 \\
& 6240.710  &  2.22  &  -3.32 \\
& 6265.140  &  2.18  &  -2.57 \\
& 6380.740  &  4.19  &  -1.32 \\
& 6392.610  &  2.28  &  -4.01 \\
& 6499.010  &  0.96  &  -4.62 \\
& 6581.220  &  1.48  &  -4.66 \\
& 6591.380  &  4.59  &  -1.98 \\
& 6608.100  &  2.28  &  -4.01 \\
& 6627.620  &  4.55  &  -1.44 \\
& 6653.920  &  4.15  &  -2.41 \\
& 6703.580  &  2.76  &  -3.01 \\
& 6710.300  &  1.48  &  -4.80 \\
& 6725.430  &  4.10  &  -2.18 \\
& 6726.740  &  4.61  &  -1.04 \\
& 6733.150  &  4.64  &  -1.45 \\
& 6739.520  &  1.56  &  -4.90 \\
& 6750.160  &  2.42  &  -2.62 \\
& 6752.710  &  4.64  &  -1.20 \\
& 6786.930  &  4.19  &  -1.95 \\
& 6820.370  &  4.64  &  -1.17 \\
& 6839.840  &  2.56  &  -3.36 \\
& 6855.720  &  4.61  &  -1.73 \\
& 6861.950  &  2.42  &  -3.80 \\
& 6862.500  &  4.56  &  -1.35 \\
& 5806.730  &  4.61  &  -0.90 \\
Fe II..........
& 5425.260  &  3.20  &  -3.18 \\
& 5991.380  &  3.15  &  -3.48 \\
& 6084.170  &  3.20  &  -3.75 \\
& 6149.250  &  3.89  &  -2.70 \\
& 6247.560  &  3.89  &  -2.30 \\
& 6369.450  &  2.89  &  -4.11 \\
Ni~I (6.25)....
& 6767.780  &  1.83  &  -2.09 \\
\enddata



\end{deluxetable}


\begin{deluxetable}{rllrrrrr}
\tabletypesize{\small}
\tablecaption{PHOTOSPHERE PARAMETERS FROM SPECTROSCOPY\label{table.stars}}
\tablewidth{0pt}
\tablehead{
\colhead{HD} & \colhead{S.T.\tablenotemark{a}} & \colhead{$T_{\rm e}$} & \colhead{$\log~g$} & \colhead{$\xi_t$~(km~sec$^{-1}$)} & \colhead{[Fe/H]} & \colhead{$n_{{\rm Fe I}}$} & \colhead{$n_{{\rm Fe II}}$}
}
\startdata
166 & K0 & 5620(40) & 4.62(0.07) & 1.60(0.11) & 0.10(0.04) & 44 & 7\\
1237\tablenotemark{b} & G6 & 5502(48) & 4.47(0.08) & 1.20(0.09) & 0.14(0.04) & \nodata & \nodata\\
1835 & G3 & 5675(60) & 4.31(0.12) & 1.30(0.13) & 0.17(0.05) & 44 & 5 \\
7590 & G0 & 5940(75) & 4.50(0.14) & 1.50(0.23) & -0.03(0.06) & 39 & 6\\
10008 & G5 & 5415(50) & 4.63(0.08) & 1.55(0.14) & 0.02(0.04) & 45 & 5\\
10780 & K0 & 5295(40) & 4.27(0.08) & 1.00(0.07) & -0.02(0.03) & 46 & 5\\
11131 & G0 & 5700(60) & 4.21(0.15) & 1.20(0.13) & -0.09(0.05) & 42 & 5\\
20630 & G5 & 5747(49) & 4.53(0.06) & 1.15(0.09) & 0.11(0.04) & 46 & 6\\
26923 & G0 & 5995(68) & 4.52(0.05) & 1.20(0.17) & 0.05(0.06) & 42 & 5\\
30495 & G3 & 5791(37) & 4.42(0.06) & 1.33(0.09) & -0.02(0.03) & 42 & 5\\
36435\tablenotemark{c} & G8 & 5510 & 4.78 & 1.15 & 0.03 & \nodata & \nodata \\
37394 & K1 & 5295(47) & 4.42(0.12) & 1.30(0.09) & 0.08(0.04) & 44 & 6\\
41593 & K0 & 5296(66) & 4.40(0.12) & 1.10(0.12) & 0.07(0.04) & 47 & 4\\
43162 & G5 & 5473(56) & 4.10(0.09) & 1.35(0.14) & -0.11(0.04) & 40 & 6\\
52698 & K1 & 5242(61) & 4.61(0.14) & 1.20(0.11) & 0.12(0.05) & 46 & 6\\
59967 & G4 & 5902(57) & 4.66(0.09) & 1.30(0.14) & 0.01(0.05) & 45 & 5\\
63433 & G5 & 5763(74) & 4.44(0.10) & 1.60(0.17) & 0.02(0.06) & 39 & 4\\
72760 & G5 & 5332(53) & 4.61(0.10) & 1.30(0.10) & 0.01(0.04) & 45 & 5\\
72905 & G1.5 & 5850(70) & 4.46(0.08) & 1.50(0.16) & -0.03(0.06) & 47 & 6\\
73350 & G0 & 5743(43) & 4.23(0.07) & 1.18(0.08) & 0.08(0.04) & 44 & 5\\
82443 & K0 & \nodata & \nodata & \nodata & \nodata  & \nodata & \nodata \\
97334 & G0 & 5736(75) & 4.14(0.07) & 1.00(0.15) & 0.05(0.06) & 45 & 6\\
109011 & K2 & 5207(72) & 4.48(0.12) & 1.30(0.14) & -0.20(0.06) & 38 & 6 \\
113449 & G6 & 5287(78) & 4.71(0.11) & 1.30(0.15) & -0.06(0.05) & 44 & 5\\
116956 & G9 & 5380(52) & 4.56(0.07) & 1.35(0.11) & 0.11(0.04) & 45 & 5\\
118972 & K2 & 5256(48) & 4.10(0.20) & 1.30(0.09) & -0.05(0.04) & 42 & 5\\
128400 & G5 & \nodata & \nodata & \nodata & \nodata  & \nodata & \nodata \\
128987 & G6 & 5539(51) & 4.49(0.06) & 1.10(0.12) & 0.00(0.04) & 46 & 5\\
130948 & G1 & 5832(50) & 4.16(0.07) & 1.05(0.11) & -0.04(0.04) & 42 & 5\\
135599 & K0 & 5300(48) & 4.53(0.05) & 1.25(0.10) & -0.06(0.04) & 44 & 5\\
141272 & G8 & 5425(51) & 4.77(0.08) & 1.50(0.11) & 0.00(0.04) & 43 & 6 \\
152391 & G8 & 5418(51) & 4.45(0.10) & 1.00(0.09) & -0.02(0.04) & 45 & 5\\
165185 & G5 & 5681(77) & 4.02(0.11) & 1.15(0.17) & -0.14(0.06) & 38 & 6\\
180161 & G8 & 5464(58) & 4.53(0.09) & 1.20(0.09) & 0.15(0.04) & 44 & 5 \\
203244 & G5 & \nodata & \nodata & \nodata & \nodata & \nodata & \nodata \\
206860 & G0 & 5900(103) & 4.35(0.10) & 1.25(0.28) & -0.02(0.08) & 40 & 5\\
217813 & G5 & 5825(50) & 4.47(0.10) & 1.05(0.11) & 0.07(0.04) & 41 & 6 \\
220182 & K1 & 5279(69) & 4.54(0.09) & 1.30(0.12) & 0.02(0.05) & 37  & 4 \\
\enddata


\tablenotetext{a}{From SIMBAD}
\tablenotetext{b}{From Gonzalez et al.~(2001)}
\tablenotetext{c}{From Santos et al.~(2001)}

\end{deluxetable}

\begin{deluxetable}{rllrr}
  \tabletypesize{\small} \tablecaption{PHOTOSPHERE PARAMETERS FROM PHOTOMETRY \label{table.irtm}} 
\tablewidth{0pt} \tablehead{ \colhead{HD} & \colhead{$T_{\rm e}$} & \colhead{$R/R_{\odot}$} & \colhead{N} & \colhead{$\chi^2$}}
\startdata
  166 &  \nodata & \nodata & \nodata & \nodata \\
  1237 & \nodata & \nodata & \nodata & \nodata \\
   1835 & 5792(90) & 0.951(0.020) & 15 & 26.9\\
  7590 & 5803(80) & 0.990(0.011) & 2 & \nodata\\
  10008 & 5184(38) & 0.810(0.009) & 4 & 8.9\\
  10780 & 5303(35) & 0.818(0.007) & 4 & 5.3\\
  11131 & 5629(45) & 0.986(0.011) & 4 & 13.5\\
  20630 & 5614(85) & 0.940(0.021) & 14 & 13.3\\
  26923 & \nodata & \nodata & \nodata & \nodata \\
  30495 & \nodata & \nodata & \nodata & \nodata \\
  36435 & 5277(28) & 0.871(0.007) & 2 & \nodata \\
  37394 & 5149(57) & 0.834(0.019) & 5 & 1.9\\
  41593 & 5120(65) & 0.830(0.013) & 2 & \nodata \\
  43162 & 5472(31) & 0.901(0.008) & 2 & \nodata \\
  52698 & 4960(25) & 0.877(0.008) & 2 & \nodata \\
  59967 & 5666(83) & 0.944(0.019) & 14 & 17.5\\
  63433 & 5500(44) & 0.910(0.010) & 4 & 8.7\\
  72760 & 5176(82) & 0.870(0.012) & 4 & 5.5\\
  72905 & 5784(86) & 0.946(0.014) & 5 & 2.1\\
  73350 & 5659(103) & 0.991(0.027) & 2 & \nodata\\
  82443 & 5142(32) & 0.839(0.008) & 3 & 0.13 \\
  97334 & 5738(35) & 1.025(0.010) & 2 & \nodata \\
  109011 & 5000(79) & 0.850(0.011) & 3 & 32.5\tablenotemark{a}\\
  113449 & 5013(139) & 0.865(0.027) & 3 & 6.6\\
  116956 & 5193(45) & 0.885(0.015) & 4 & 4.0\\
  118972 & \nodata & \nodata & \nodata & \nodata \\
  128400 & \nodata & \nodata & \nodata & \nodata \\
  128987 & 5326(39) & 0.916(0.009) & 4 & 11.7\\
  130948 & 5813(55) & 1.058(0.014) & 5 & 6.9\\
  135599 & \nodata & \nodata & \nodata & \nodata \\
  141272 & \nodata & \nodata & \nodata & \nodata \\
  152391 & \nodata & \nodata & \nodata & \nodata \\
  165185 & 5863(70) & 0.968(0.013) & 10 & 10.0\\
  180161 & 5342(30) & 0.844(0.008) & 2 & \nodata \\
  203244 & \nodata & \nodata & \nodata & \nodata \\
  206860 & \nodata & \nodata & \nodata & \nodata \\
  217813 & 5644(35) & 1.068(0.010) & 2 & \nodata \\
  220182 & 5208(35) & 0.854(0.008) & 4 & 46.1\\
  \enddata


\tablenotetext{a}{Possible spectroscopic binary}

\end{deluxetable}

\begin{deluxetable}{rrrrrrrrrr}
\tabletypesize{\scriptsize}
\tablecaption{DERIVED ELEMENTAL ABUNDANCES \label{table.abundance}}
\tablewidth{0pt}
\tablehead{
\colhead{HD} & \colhead{[Fe/H]} & \colhead{[C/H]}& \colhead{[Na/H]}& \colhead{[Mg/H]}& \colhead{[Si/H]}& \colhead{[S/H]}& \colhead{[Ca/H]}& \colhead{[Ti/H]}& \colhead{[Ni/H]}
}
\startdata
166  &  0.10(0.04) & -0.27(0.07) & 0.09(0.04) & -0.03(0.07) & 0.06(0.03) & -0.01(0.07) & 0.09(0.08) & 0.13(0.05) & -0.04(0.08)\\
1237  &  0.16(0.03) & \nodata & 0.08(0.03) & 0.09(0.05) & 0.16(0.01) & 0.24(0.05) & 0.19(0.03) & 0.11(0.03) & 0.10(0.06)\\
1835  &  0.17(0.05) & 0.06(0.10) & 0.01(0.07) & 0.10(0.10) & 0.26(0.02) & 0.37(0.09) & 0.21(0.07) & 0.12(0.07) & 0.17(0.12)\\
7590  &  -0.03(0.06) & -0.17(0.12) & -0.22(0.07) & -0.14(0.12) & 0.07(0.09) & -0.09(0.11) & -0.04(0.06) & -0.16(0.08) & -0.16(0.14)\\
10008  &  0.02(0.04) & \nodata & -0.16(0.04) & -0.09(0.10) & -0.02(0.04) & 0.30(0.10) & 0.01(0.05) & 0.02(0.07) & -0.21(0.11)\\
10780  &  -0.02(0.03) & -0.03(0.07) & -0.09(0.03) & -0.02(0.07) & 0.02(0.02) & \nodata & 0.03(0.04) & -0.02(0.05) & -0.15(0.08)\\
11131  &  -0.09(0.05) & -0.12(0.10) & -0.23(0.04) & -0.17(0.10) & -0.01(0.02) & \nodata & -0.06(0.07) & -0.15(0.06) & -0.37(0.10)\\
20630  &  0.11(0.04) & -0.04(0.11) & -0.04(0.03) & -0.04(0.08) & 0.07(0.04) & \nodata & 0.10(0.04) & 0.11(0.08) & 0.04(0.10)\\
26923  &  0.05(0.06) & -0.03(0.12) & -0.11(0.10) & -0.03(0.10) & 0.02(0.06) & -0.06(0.10) & -0.07(0.35) & 0.02(0.12) & -0.08(0.11)\\
30495  &  -0.02(0.03) & -0.10(0.06) & -0.09(0.05) & -0.13(0.06) & 0.02(0.02) & -0.07(0.06) & -0.01(0.03) & 0.01(0.08) & -0.26(0.07)\\
37394  &  0.08(0.04) & -0.09(0.09) & 0.03(0.08) & 0.00(0.09) & 0.08(0.03) & 0.00(0.09) & 0.13(0.05) & 0.09(0.07) & -0.05(0.10)\\
41593  &  0.07(0.04) & \nodata & -0.14(0.06) & 0.02(0.10) & 0.11(0.06) & 0.10(0.10) & 0.06(0.06) & 0.14(0.09) & 0.01(0.11)\\
43162  &  -0.11(0.04) & -0.08(0.10) & -0.05(0.09) & -0.03(0.10) & -0.04(0.04) & -0.04(0.09) & -0.03(0.05) & -0.10(0.08) & -0.26(0.11)\\
52698  &  0.12(0.05) & 0.48(0.11) & 0.04(0.13) & 0.07(0.11) & 0.14(0.05) & 0.04(0.11) & 0.13(0.06) & 0.31(0.10) & 0.01(0.12)\\
59967  &  0.01(0.05) & -0.01(0.12) & -0.14(0.08) & -0.12(0.09) & -0.04(0.05) & 0.20(0.09) & 0.00(0.04) & 0.07(0.09) & -0.12(0.11)\\
63433  &  0.02(0.06) & -0.10(0.11) & -0.10(0.12) & -0.02(0.12) & 0.10(0.04) & -0.15(0.11) & 0.12(0.09) & 0.08(0.08) & -0.07(0.14)\\
72760  &  0.01(0.04) & \nodata & 0.02(0.18) & -0.13(0.09) & 0.09(0.02) & 0.27(0.09) & 0.06(0.05) & 0.00(0.07) & -0.14(0.10)\\
72905  &  -0.03(0.06) & -0.07(0.10) & -0.09(0.04) & -0.13(0.11) & -0.05(0.03) & \nodata & -0.03(0.05) & -0.15(0.06) & -0.37(0.10)\\
73350  &  0.08(0.04) & -0.05(0.07) & -0.05(0.08) & 0.04(0.07) & 0.08(0.01) & -0.08(0.07) & 0.10(0.04) & 0.01(0.05) & -0.02(0.08)\\
97334  &  0.05(0.06) & 0.03(0.05) & -0.03(0.04) & 0.01(0.12) & 0.02(0.02) & 0.08(0.11) & 0.06(0.07) & -0.03(0.07) & 0.05(0.14)\\
109011  &  -0.20(0.06) & -0.05(0.11) & -0.29(0.09) & -0.28(0.11) & -0.19(0.09) & \nodata & -0.09(0.09) & -0.03(0.13) & -0.55(0.12)\\
113449  &  -0.06(0.05) & 0.36(0.12) & -0.22(0.07) & -0.17(0.12) & -0.05(0.08) & \nodata & -0.01(0.08) & -0.02(0.10) & -0.11(0.14)\\
116956  &  0.11(0.04) & 0.14(0.09) & -0.04(0.12) & -0.07(0.09) & 0.13(0.05) & \nodata & 0.08(0.05) & 0.08(0.07) & 0.07(0.11)\\
118972  &  -0.05(0.04) & 0.36(0.10) & -0.13(0.12) & 0.12(0.10) & -0.09(0.03) & \nodata & 0.09(0.10) & 0.02(0.07) & -0.24(0.09)\\
128987  &  0.00(0.04) & 0.26(0.08) & -0.12(0.04) & -0.08(0.08) & 0.00(0.02) & \nodata & 0.01(0.05) & -0.01(0.06) & -0.08(0.10)\\
130948  &  -0.04(0.04) & -0.10(0.04) & -0.20(0.08) & -0.07(0.07) & 0.02(0.03) & -0.03(0.07) & -0.03(0.04) & -0.19(0.05) & -0.16(0.09)\\
135599  &  -0.06(0.04) & 0.50(0.71) & -0.23(0.08) & -1.27(0.00) & -0.05(0.05) & \nodata & -0.05(0.04) & -0.01(0.06) & -0.19(0.09)\\
141272  &  0.00(0.04) & \nodata & -0.22(0.05) & -0.18(0.08) & 0.00(0.07) & \nodata & -0.02(0.05) & 0.03(0.07) & -0.17(0.09)\\
152391  &  -0.02(0.04) & -0.31(0.10) & -0.17(0.06) & -0.09(0.09) & -0.06(0.03) & -0.14(0.09) & -0.03(0.05) & -0.01(0.06) & -0.09(0.10)\\
165185  &  -0.14(0.06) & -0.13(0.12) & -0.34(0.08) & -0.08(0.12) & -0.10(0.04) & -0.46(0.11) & -0.05(0.06) & -0.31(0.08) & -0.39(0.13)\\
180161  &  0.15(0.04) & -0.13(0.09) & 0.00(0.06) & 0.05(0.09) & 0.14(0.05) & 0.08(0.09) & 0.15(0.05) & 0.15(0.07) & 0.06(0.10)\\
206860  &  -0.02(0.08) & -0.26(0.24) & -0.16(0.06) & -0.13(0.16) & -0.05(0.03) & -0.08(0.14) & 0.00(0.08) & -0.07(0.12) & -0.30(0.18)\\
217813  &  0.07(0.04) & -0.38(0.07) & -0.06(0.03) & -0.08(0.08) & 0.04(0.02) & -0.23(0.07) & 0.03(0.04) & -0.03(0.06) & -0.10(0.09)\\
220182  &  0.02(0.05) & 1.38(0.10) & -0.17(0.07) & -0.11(0.10) & 0.07(0.06) & \nodata & 0.08(0.06) & -0.04(0.10) & -0.14(0.11)\\
\enddata
\end{deluxetable}






\end{document}